# Optical-mechanical operation of the F2T2 Filter: a tunable filter designed to search for *First Light*

Erin Mentuch[a,b], Alan Scott[b,c], Roberto Abraham[a], Elizabeth Barton[d], Matthew Bershady[e], Joss Bland-Hawthorn[f], David Crampton[g], René Doyon[h], Steve Eikenberry[i], Mike Gladders[j], Karl Glazebrook[k], Joe Jenson[i], Jeff Julian[i], Roger Julian[i], Jean-Paul Kneib[l], David Loop[g], Nick Raines[i], Neil Rowlands[b], and JD Smith[m]

[a]Department of Astronomy & Astrophysics, University of Toronto
[b]COM DEV Canada
[c]Centre for Research in Earth & Space Science, York University
[d]Department of Physics & Astonomy, University of California, Irvine
[e]Department of Astronomy, University of Wisconsin
[f]Anglo-Australian Observatory, Australia
[g]Herzberg Institute of Astrophysics, National Research Council Canada
[h]Departement de Physique, Université de Montreal
[i]Department of Astronomy, University of Florida
[j]Department of Astronomy and Astrophysics, University of Chicago
[k]Centre for Astrophysics & Supercomputing, Swinburne University of Technology
[l]Laboratoire d'Astrophysique de Marseilles
[m]Ritter Astrophysical Observatory, University of Toledo

## ABSTRACT

The Flamingos-2 Tandem Tunable filter is a tunable, narrow-band filter, consisting of two Fabry-Perot etalons in series, capable of scanning to any wavelength from 0.95 to 1.35 microns with a spectral resolution of R~800. It is an accessory mode instrument for the near-IR Flamingos-2 imaging-spectrograph designed for the Gemini South 8m Observatory and will be fed through the upcoming Multi-Conjugate Adaptive Optics feed. The primary science goal of the F2T2 filter is to perform a ground-based search for the first star forming regions in the universe at redshifts of $7 < z < 11$. The construction of the F2T2 filter is complete and it is currently in its calibration and commissioning phases. In this proceeding, we describe the calibration and performance of the instrument.

## 1. INTRODUCTION

The Flamingos-2 Tandem Tunable filter[1] (F2T2) is an accessory mode instrument designed for the Flamingos-2[2] near-IR wide field imager and multi-object spectrometer, built by the University of Florida, for use on the Gemini South 8m telescope. F2T2 is a 60 mm-clear aperture, IR-optimized, air spaced, tandem-etalon tunable Fabry-Perot filter, which will investigate the first sources of star formation in the Universe. The optical components and electronics for the device were designed and assembled by COM DEV Canada. Construction of the device has been funded by a Steacie Fellowship and a Canada Foundation for Innovation Career Award given to the project's Canadian-PI (R. Abraham).

In many aspects, the F2T2 filter is a ground-based analog to the Tunable Filter Imager[3] (TFI) presently being constructed as a component of the James Webb Space Telescope's (JWST) Fine Guidance Sensor Camera. JWST's TFI and Gemini's F2T2 share key optics and electronics and like F2T2, the primary science goal of TFI is to probe star formation in the distance Universe. However, unlike F2T2, TFI is designed to withstand the challenges of space launch and flight. In addition, F2T2 is a dual etalon device operating at higher resolution than TFI, a single etalon device. The second etalon suppresses the wings of the airy transmission function, allowing F2T2 to operate with high efficiency in between contaminating OH lines, which dominate the near-IR sky background emission redward of 0.9 μm.

|  | Tunable Filter Imager (TFI) | Flamgingos-2 Tandem Tunable Filter (F2T2) |
|---|---|---|
| Location in Telescope | Near the pupil of James Webb Space Telescope (JWST) | Inside Flamingos-2 imaging spectrograph near the focus of Gemini South 8m Observatory |
| Spectral Resolution | 100 | > 800 (wing suppressed) |
| Wavelength Range | 1.5 μm – 5 μm | 0.95 μm – 1.35 μm |
| Field of View | ~ 2.2′ | ~ 50″ (with Multi-Conjugate Adaptive Optics (MCAO)) <br> ~ 1.5′ (without MCAO) |
| Image Quality | Diffraction Limited | MCAO |
| Operating Temperature | ~35 K | ~110 K |
| Principal Investigator | R. Doyon (Univ. of Montreal) | R. Abraham (Univ. of Toronto) |
| First Light | 2013+ | Summer 2008 (1.6m Mont-Mégantic Observatory) <br> Early 2009 (8m Gemini South Observatory) |

**Table 1**: Key differences between the JWST's TFI and Gemini's F2T2.

### 1.1 Science Motivation

The formation of the first luminous, self-gravitating objects marks a critical transition of the Universe from its smooth initial state to its clumpy current state. It is believed that these objects, topically referred to as *First Light*, probably triggered cosmic reionization, the critical transition from a Universe dominated by neutral gas, which fragments on small scales, typically the size of globular clusters, to one dominated by ionized gas, which fragments on much larger scales and leads to the formation of galaxy-sized objects. Recent analysis of Five-year Wilkinson Microwave Anisotropy Probe (WMAP) temperature and polarization measurements suggests a lower limit on reionization of $z_{reion}>6.7$ with 3σ confidence[4]. In agreement, temperature measurements of the intergalactic medium (IGM) at $z=2-4$ suggest that reionization occurred at $z<9$[5] and spectral signatures of a number of distant quasars indicate that reionization was over by $z\sim6.5$[6].

The nature of the sources of *First Light* is also constrained by current observations. Galaxies, with confirmed spectroscopic redshifts, have been detected out to $z\sim7$[7]. Two distinct classes of very high redshift objects have been found using two (biased) selection techniques. Lyman break galaxies (LBGs) are found using the Lyman dropout technique, which looks for the redshifted signature of the UV continuum break between 91.2 nm and 121.6 nm. The relative efficiency in telescope time to detect LBGs has allowed for spectroscopic confirmation of about 1000 of these objects at $z\sim3-6$[8]. Over 500 more have been confidently detected at $z\sim6-7$, providing very high redshift measurements of the star formation history (SFH) and mass density[9,10]. However, only the most massive of these galaxies can be found with this selection technique at such high redshifts with today's telescopes. Studies of candidates at $z\sim7$ indicate that the number density of these most luminous galaxies was not sufficient to reionize the universe[11] and thus points to less massive star-forming objects providing the bulk of the ionizing radiation to reionize the universe.

Lyα emitters (LAEs), the other class of high-z galaxies, are star-forming galaxies with strong Lyα (λ=121.6 nm) emission detected through wide-field, narrow-band surveys[12]. This technique can detect less massive galaxies, as the line emission flux can be 10-300 times the stellar continuum flux. Currently, more than 100 LAE candidates have been found at $z\sim6.5$, including more than 30 with spectroscopic confirmation[13]. Because of the small volumes probed with standard narrowband filters, cosmic variance leads to large uncertainties in the number counts of LAEs. As a result, it is still questionable if these relatively low-mass but abundant, star-forming galaxies produce enough ionization radiation to fully ionize the universe. Measuring the star formation rates and mass density of the most distant objects will provide the missing pieces to our understanding of the epoch of reionization and the sources of *First Light* responsible for it.

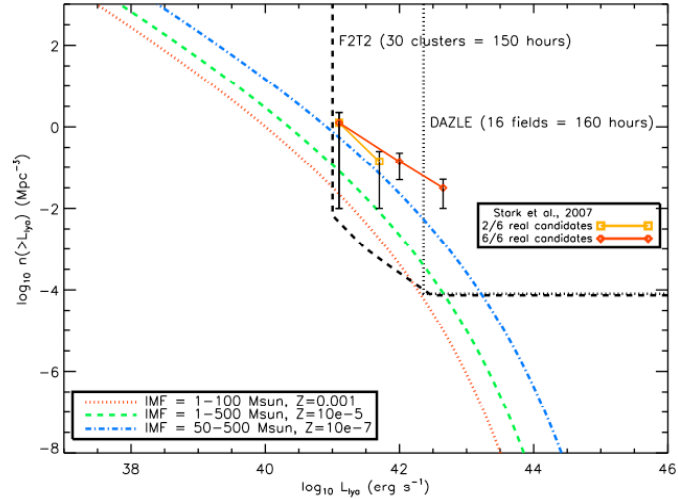

**Figure 1.** The luminosity function of LAEs calculated from the initial dark matter halo mass function. A simple prescription translating baryonic density to ionizing radiation re-emitted as Lyman alpha emission at 121.6 nm is used. The three curves are represented by different IMF and metallicity scenarios. The figure shows the survey limits of a mock Gemini Genesis Survey taken with F2T2 for 30 galaxy clusters in a time frame of 150 hours. It also shows the survey limits of a current search for high-z LAEs undertaken with the DAZLE instrument consisting of 16 wide-field observations with four narrowband filters.

Finding sources of *First Light* beyond z>7 is difficult to do with both broad-band filters and standard narrowband filters. The dropout technique at this redshift requires deep observations in the NIR, but beyond 0.9 μm, the night sky is contaminated with a number of narrow emission lines from atmospheric hydroxyl molecules. Because of this high sky noise, it is just not feasible to reach the low mass galaxies from the ground with today's 8-10m class observatories. It won't be until the next generation of observatories, such as the James Webb Space Telescope (JWST), scheduled for launch in 2013, that less massive galaxies can be studied using this selection technique. Multiple narrowband searches, high in fabrication costs, can potentially find LAEs before JWST, but require a lot of observing time in order to probe volumes large enough to limit uncertainties due to cosmic variance. To achieve this, a narrow-band imaging instrument has been designed for the 8.2m Very Large Telescope (VLT) in Chile, called the Dark Age z (redshift) Lyα Explorer[14] (DA*z*LE). DA*z*LE consists of a pair of interchangeable narrow-band filters at four wavelengths in the NIR, covering 6.5<z<12. DA*z*LE can detect LAE flux sensitivities of $\sim 1.5 \times 10^{42}$ erg s$^{-1}$ and covers a survey volume of 6900 Mpc$^{-3}$. Now operating at VLT, DA*z*LE has been successful at detecting a single z=7.7 LAE candidate[1], which still requires spectroscopic follow-up.

Aside from traditional narrow-band imaging surveys, there are a few novel ways in which astronomers are hoping to find sources of *First Light* prior to JWST. Assuming the sources are faint, then a better approach is to take advantage of the magnification due to the gravitational lensing from intermediate redshift galaxy clusters to brighten the apparent magnitude of a low luminosity, high-z galaxy. Such searches have already been performed using standard slit spectroscopy and have been successful at detecting a number of LAEs at 4.5<z<5.6[15,16]. From these surveys, it is evident that the number density of massive objects is much lower at high redshifts, consistent with the standard Hierarchical model of galaxy formation. Other searches have found likely candidates at z > 8, but spectroscopic confirmation of their redshift is still needed[17,18]. If the candidates in these searches are real, the number density (plotted in Figure 1 above) implies that low-luminosity, star-forming systems are more abundant at z>9 than at z<6.

Slit searches probe very small spatial volumes, but have high coverage in redshift (temporal) space, while traditional narrowband filter searches can probe large spatial volumes, but not a lot of redshift space. A novel, relatively low budget approach is to perform a search with an instrument that can probe large volumes in both spatial and redshift space. The Gemini Genesis Survey (GGS) will expand the methodology of slit-based lensing LAE search surveys by utilizing the tunable, narrow-band F2T2 filter to image the entire plane of a strong intermediate redshift galaxy cluster, covering more volume than a spectroscopic slit and much more redshift space than a single narrowband filter.

---

[1] http://www.ast.cam.ac.uk/~optics/dazle/VLT_2006/DAZLE_results/z7_index.html

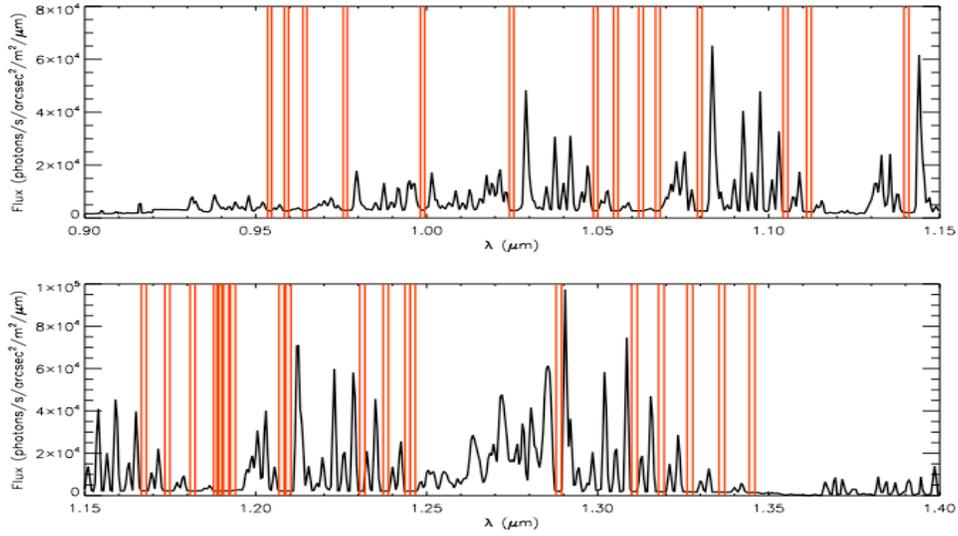

**Figure 2.** Night sky spectrum in the NIR. The vertical red lines indicate 40 possible wavelength slices to probe with the F2T2 narrowband filter. For wide spectral windows, multiple slices can be potentially probed. We avoid probing the sky where $H_2O$ (at 0.95, 1.13 and 1.41 μm) and $O_2$ (at 1.27 μm) molecular absorption bands are found.

**1.2 Design of the Gemini Genesis Survey**

The primary science goal of the F2T2 filter is to search for high-redshift Lyα emission in young star-forming systems. At $z>7$, the Lyα emission line at 121.6 nm is redshifted to NIR wavelengths longer than 0.9 μm. Beyond this wavelength, the night sky is increasingly bright in the NIR, about 20-50 times brighter than the optical night sky, plagued with strong hydroxyl emission lines from molecules in the atmosphere. F2T2 has been optimized to search between these narrow emission lines where the sky is much fainter (about 2000 photons/s/m$^2$/arcsec$^2$/μm). This fainter sky noise allows F2T2 to reach a 5σ flux sensitivity of ~3-6×10$^{-18}$ erg s$^{-1}$ cm$^{-2}$. Figure 2 presents 40 possible wavelength regions that F2T2 can probe, keeping in mind that we must also avoid regions of atmospheric molecular observation of $H_2O$ at 0.95, 1.13 and 1.41 μm and $O_2$ at 1.27 μm. The resolution of each slice is $\lambda/\Delta\lambda = 800$ (about 0.0012-0.0019 μm). With MCAO, the 45" field of view of F2T2 can probe a comoving volume of ~11 Mpc$^3$ for each wavelength slice. Without MCAO, the field of view of F2T2 is 1.'5 and can probe a comoving volume of ~44 Mpc$^3$. For a single field of view observed at 40 wavelength slices, we are capable of probing total volumes of ~440 Mpc$^3$ and ~1800 Mpc$^3$ with and without F2T2 being fed through the MCAO on Gemini South.

Using a simple star formation model[19], a Lyα luminosity function can be predicted based on the halo mass function and how many ionizing photons are re-emitted in the Lyα line. Figure 1 shows a theoretical Lyα luminosity function based on this simple prescription with three different scenarios of metallicity and the initial mass function (IMF). The nature of the IMF directly affects how many ionizing photons a galaxy will emit, with heavier IMFs, which contain more massive stars, emitting much more ionizing radiation, which will result in a brighter Lyα luminosity function than an IMF with a lower fraction of massive stars[20]. Overplotted on Figure 1 are the predicted volumes and fluxes reached for two proposed surveys, one with the F2T2 filter (i.e. GGS) and one with DA$z$LE. The GGS will probe 30 galaxy clusters at 40 different wavebands, while DAZLE will probe 16 fields with 4 different filters, with both surveys requiring similar exposure times of about 150 hours. It is apparent in Figure 1, that F2T2 will reach greater flux limits by taking advantage of the magnification gain in flux provided by gravitationally lensing galaxy clusters. For reference, the Lyα luminosity function measured from objects detected in a spectroscopic slit lensing survey[21] are shown in red and orange. If all of the candidates in this survey are real (red line), then DA$z$LE and F2T2 should both expect to find objects, but if only two of the candidates are real (orange line), the GGS with F2T2 will likely be more successful.

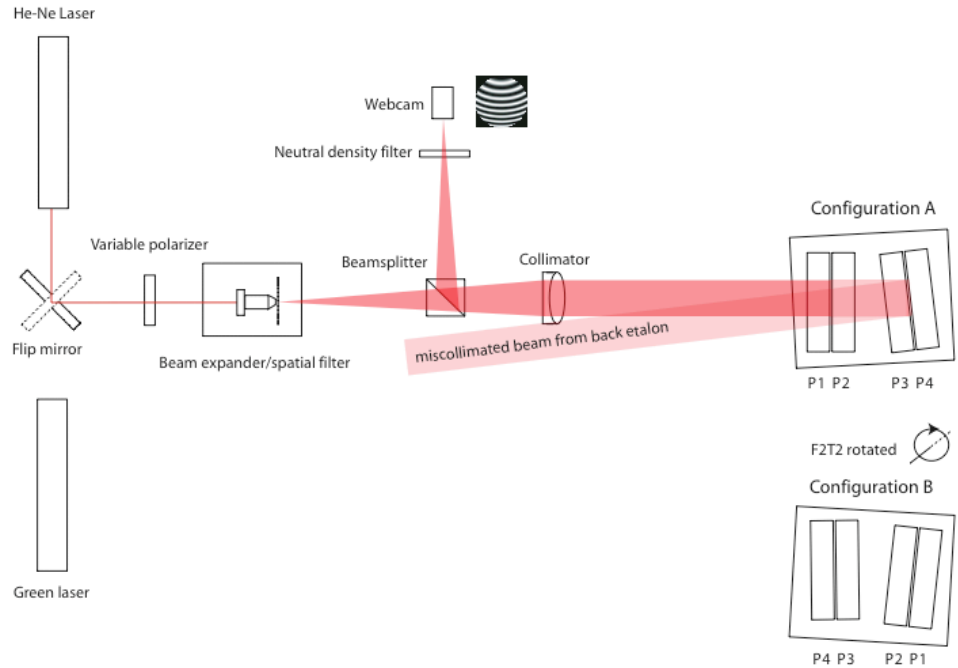

**Figure 3:** Optical setup for calibrating the F2T2 filter. The combination of two lasers allows for absolute gap determination.

Currently, complimentary surveys[2] to find the epoch of reionization are being undertaken by measuring the distribution of 21cm line emission from neutral hydrogen at extremely high redshifts. As the universe is ionized, patches of ionized hydrogen are identified out of the sea of neutral hydrogen by the lack of 21cm line emission. The GGS will be able to use the results of the current surveys in order to identify regions of ionized hydrogen and focus in on the sources that caused the reionization.

## 2. ETALON OPERATION

The F2T2 system consists of two etalons in tandem. The etalons have been polished and coated to provide a reflective finesse of >30 over the wavelength range of 0.95 μm to 1.35 μm. In order to reach the required finesse, the etalons must be controlled down to a precision of <10 nm. The system is controlled using two independently commanded Multi-Application Low Voltage Piezoelectric Instrument Control Electronic[22] (MALICE) rack-mounted systems. The MALICE system drives low voltage piezoelectric actuators (LVPZTs) using capacitive displacement sensors (CDSs) positional feedback technology. The control software has been developed by COM DEV Canada in C language on an Analog DSP 2191M EZKIT evaluation board using Analog Devices VisualDSP++ environment. There are 6 LVPZTs and 5 CDSs per etalon, comprising 3 capacitive bridges to provide high accuracy feedback on the gap and wedge. Two pairs are used to provide 'X' and 'Y' parallelism, while a 5th sensor is used in a 'Z' bridge with a fixed ceramic capacitor to provide an absolute gap reference.

---

[2] Giant Metrewave Radio Telescope (GMRT; `http://www.ncra.tifr.res.in`), LowFrequency Array (LOFAR; `http://www.lofar.org`), Murchison Widefield Array (MWA; `http://web.haystack.mit.edu/arrays/MWA`), Primeval Structure Telescope/21CMA (PAST; `http://web.phys.cmu.edu/~past/`), and Square Kilometre Array (SKA; `http://www.skatelescope.org`).

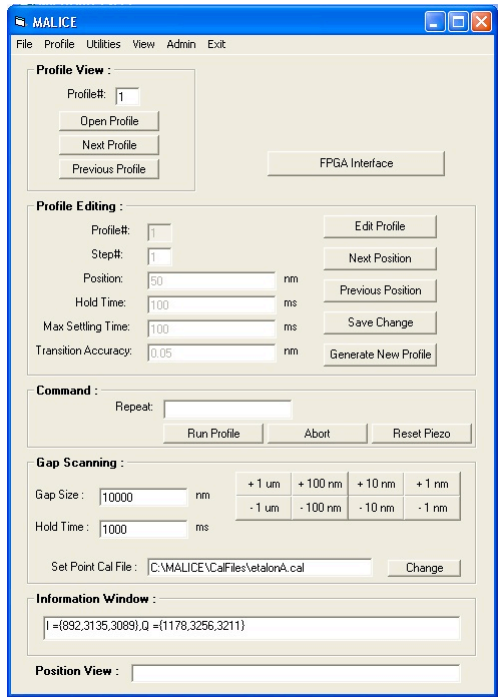

**Figure 4:** The MALICE GUI used to control each etalon. Calibration of parallel setpoints is performed under the utilities drop-down menu. Tandem alignment is performed via the 'Gap Sanning' feature. Bash profile files can be added under the 'Command' portion for automatic operation of the two etalons.

### 2.1 Calibration routine

To calibrate the etalons, parallel plate positions are set by driving the LVPZTs with external power supplies and optical feedback. Each etalon is calibrated separately in reflection using a Fizeau interferometer optical system as described in Figure 3. A single laser (in our case a HeNe at 632.8 nm) can be used to parallelize the plates, or two lasers (at different wavelengths as seen in Figure 3) can be combined to achieve a measurement of the absolute gap in between the plates. Interference fringes, observed via a webcam attached with a C-mount to the beam splitter, are used to parallelize the plates in real-time with the external variable power supplies. Capacitive bridge error signals are zeroed by altering in-phase (I) and quadrature (Q) digital bridge drive setpoints via an autocalibration routine. Intermediate gap spacings can then be linearly interpolated on all three channels of each etalon. The relationship of digital setpoints to the gap size are uploaded under the 'Gap Scanning' feature of the MALICE GUI displayed in Figure 4. This allows the etalon to be digitally tuned to a defined parallel gap size.

### 2.2 Control Methodology

The MALICE software is designed to hold the plates at a parallel position for a specified length of time and accuracy. The etalon is driven to the desired parallel position using closed-loop control to minimize the capacitive bridge errors from CDS feedback. Each plate of an etalon is supported on three equally spaced PZT actuators from a common annular baseplate, shared by both etalons. This design provides constant gap spacing from ambient down to operating temperature. The MALICE circuit has three constant current outputs with 24-bit resolution. Each output drives a pair of neighbouring PZT actuators in parallel with opposite polarity—one attached to each plate. To increase the gap, top PZTs raise the top platform while bottom PZTs lower the bottom platform. This results in an increase of the gap between the sense capacitor plates resulting in a decrease in capacitance which is used to measure the displacement. The inverse is performed to decrease the gap.

The LVPZTs are driven by two cascaded 12 bit DACs, a 'dynamic' DAC and a 'coarse' DAC. The dynamic DACs provide high current drive, while the coarse DACs provide current drives a factor of 1000 times smaller. The utilization of both drive levels allows for fast, high-bandwidth stepping at high accuracy in a 2-stage control approach. First, largefeedback errors drive open loop control to a predicted position based on stored characterization data. This is done at high speed using a high current dynamic drive circuit, and thus requires a wide bandwidth. The dynamic drive with its open loop and high bandwidth allows too much noise to meet the necessary final positioning tolerances. The second stage is based on feedback from the calibrated capacitive bridge and results in small pseudo-static current adjustments to the PZTs at a very low bandwidth. The setpoint bridge drive provides a fixed frequency sine wave, and the error signal output is strongly filtered at this frequency. The resulting digital signal is synchronously demodulated to provide a very narrow error measurement bandwidth. This 2-stage approach allows a very fast step over the entire PZT range, while still coming to a high accuracy final set point with a low noise bandwidth. The setpoint I and Q drive amplitudes for each bridge are set by a pair of cascaded 12-bit DACs, giving an effective 22 bit positioning precision. Calibration accuracy of the setpoint positions is limited mainly by the optical metrology.

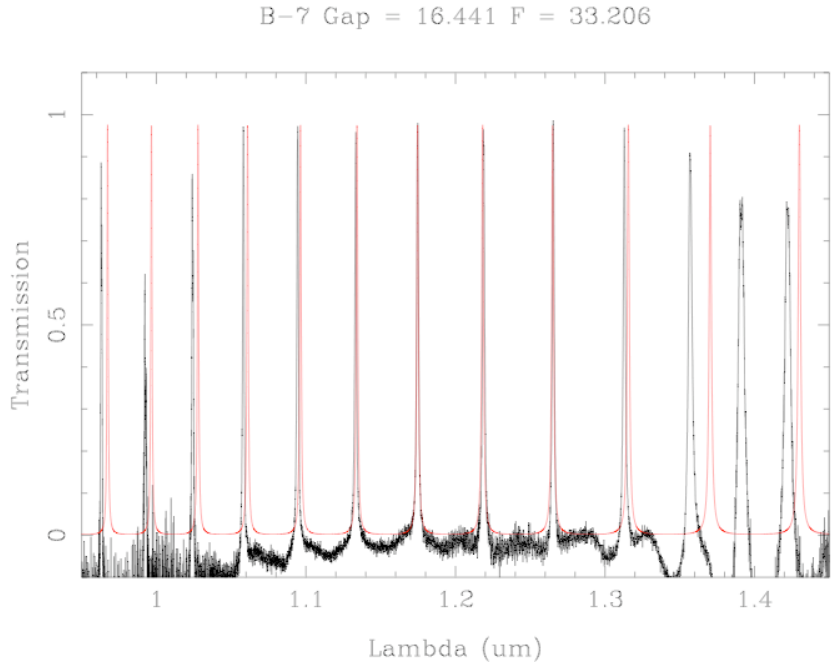

**Figure 5**: Reflection spectra of an F2T2 etalon. Fitting to a theoretical fabry-perot spectrum reveals the gap between the etalon plates is 16.4 μm and the finesse of the etalon is 33.2.

### 2.3 Tandem Operation

Co-alignment of the two etalons requires monitoring in transmission under active closed-loop control. A GUI feature, displayed in Figure 4, has been added to the MALICE system that allows the user to scan along the parallel setpoints measured from the calibration routine described in Section 2.1. Tandem alignment is performed by holding one etalon at a fixed gap spacing while scanning along parallel gap spacings of the second etalon. Maximum throughput at a desired peak wavelength can be determined using spectral feedback (ideally a fast-reading Fourier Transform Interferometer (FTIR).

Optimum free spectral range is expected to be achieved when the order separation between the two etalons, $\Delta m = 2$. It is expected to be easier to co-align the two etalons over a larger region of the surface when the order is minimized (m<20). At room temperature, the etalon gaps should scan over a range of ~20 microns from ~5 to 25 microns absolute gap. At 140 K, this range is expected to be smaller, and the response speed will be correspondingly slower. Cryostat tests are planned in Summer 2008 to investigate the effect of a colder operating temperature on tandem operation.

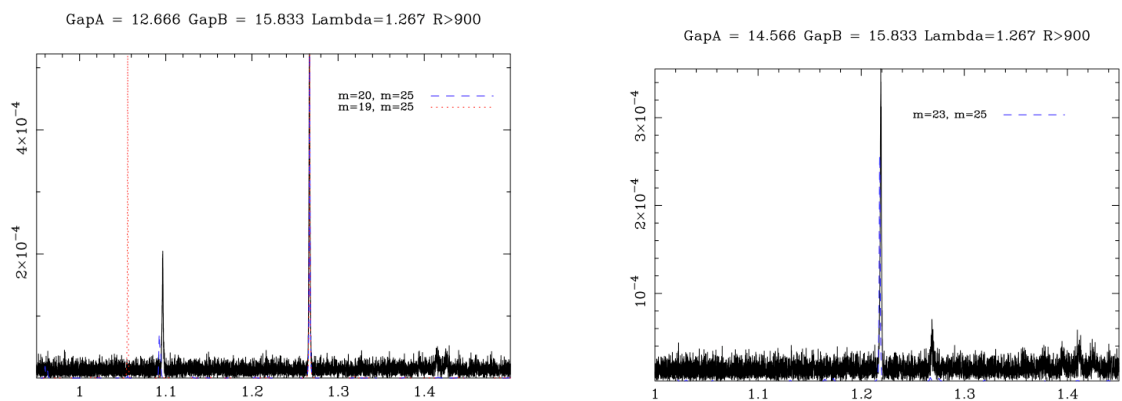

**Figure 6**: Transmission spectra of F2T2 with both etalons in tandem. In the left panel, the etalons are tuned 5 orders apart by setting the gap of Etalon A at 12.666 μm and Etalon B at 15.833 μm. Two separate tandem etalon models are plotted to show the order ratio between the two etalons. Maximum throughput at a single wavelength is achieved by tuning the etalons just two orders apart as is seen in the right panel. Etalon A has a gap of 14.566 μm and etalon B has a gap of 15.833μm.

## 3. SPECTRAL ANALYSIS

### 3.1 Parallelism

The digital setpoints, calibrated by red optical light, are verified in the near-IR by spectra taken for each etalon in reflection with a MIRMAT Fourier Transform Interferometer (FTIR). The resulting spectra are analyzed with perl-based fitting software, which fits a library of theoretical Fabry-Perot spectra at a range of finesse and absolute gap values. Figure 5 shows a reflection spectrum of one of the etalons fitted to a Fabry-Perot model with a finesse of 33.2 and a gap of 16.4 μm. The gap measured in reflection for each calibrated parallel position is used to define the relationship of CDS digital setpoints to physical gap size used to align the etalons in tandem as described in Section 2.3.

### 3.2 Tandem Operation

The relationship between absolute gap and digital setpoints can be used to align the two etalons in tandem to achieve transmission at a single wavelength by tuning the etalons to different orders at the same peak wavelength. It is found that the arrangement of a difference of two order spacings results in the highest throughput at the peak wavelength and minimizes the leakage of light at nearby orders. The use of a J-filter in the optical path of the telescope prevents light leakage at orders further away from the peak wavelength order. Figure 6 shows how tuning the etalon from an order difference of 5 to an order difference of 2 ensures the light passing through the F2T2 filter is from the desired narrow-band spectral range.

## 4. UPCOMING COMMISSIONING

The F2T2 filter is scheduled for its first commissioning run at the 1.6 m Mont-Mégantic Observatory in August 2008. The filter will be mounted inside the SIMON near-IR imager. The Flamingos-2 imaging spectrograph is expected to undergo commissioning in late 2008, during which the F2T2 filter will undergo further operational commissioning as well as some scientific commissioning at cryostat operating temperatures. The filter will be available for use with Flamingos-2 during the first general public call for scientific observations, expected in March 2009, for observations in semester 2009B.

**Acknowledgements**
The authors would like to acknowledge the support of NSERC, the Canada Foundation for Innovation, and the Canadian Space Agency. We would also like to recognize everyone on the Flamingos-2 and Gemini Observatory for their generous efforts in accommodating F2T2.